\documentclass[preprint,showpacs,superscriptaddress,preprintnumbers,amsmath,amssymb]{revtex4}

\usepackage{graphicx}
\usepackage{dcolumn}
\usepackage{bm}

\newcommand\+{\dagger}
\newcommand\diag{\operatorname{diag}}

\newcommand\Lcal{\mathcal{L}}

\newcommand\D{\mathcal{D}}

\newcommand\p{\partial}

\usepackage[normalem]{ulem}  
\usepackage[dvips]{color} 

\renewcommand\sout{\bgroup \color{red} \ULdepth=-.5ex \ULset}

\newcommand {\beq}{\begin{eqnarray}}
\newcommand {\eeq}{\end{eqnarray}}

\usepackage{amsmath}	
\begin{document}

\begin{flushright}
TKYNT-12-03
\end{flushright}

\title{
Anisotropic Optical Response of Dense Quark Matter under Rotation: 
Compact Stars as Cosmic Polarizers 
}

\author{Yuji~Hirono}
 \email{hirono@nt.phys.s.u-tokyo.ac.jp}
\affiliation{
Department of Physics, University of Tokyo, 
Hongo~7-3-1, Bunkyo-ku, Tokyo 113-0033, Japan
}

\author{Muneto Nitta}
 \email{nitta@phys-h.keio.ac.jp}
\affiliation{Department of Physics, and Research and Education Center 
for Natural Sciences, Keio University, 4-1-1 Hiyoshi, Yokohama, 
Kanagawa 223-8521, Japan}

\date{\today}

\begin{abstract}
Quantum vortices in the color-flavor locked (CFL) phase of QCD 
have bosonic degrees of freedom, called the
 orientational zero modes, localized on them.
We show that the orientational zero modes are electromagnetically charged.
As a result, a vortex in the CFL phase nontrivially interacts with
photons.
We show that a lattice of vortices acts as a polarizer 
of photons with wavelengths larger than some critical length. 
\end{abstract}

\maketitle

\emph{Introduction.}---
The strong interaction, which is one of the fundamental forces in nature, 
is fully described by quantum chromodynamics (QCD). 
QCD matter shows a rich variety of phenomena at finite
temperatures and/or baryon densities \cite{Fukushima:2010bq}, and the
determination of the phase
diagram has been a topic of considerable interest in high-energy physics.
Quark matter is expected to exhibit color superconductivity, 
triggered by quark-quark pairings,
at high baryon densities and low temperatures \cite{Alford:1998mk,Alford:2007xm}.
It has been reported in Ref.~\cite{Alford:1998mk} that 
the ground state is the color-flavor locked (CFL) phase 
at very high densities, in which the three light flavors 
(up, down and strange) of quarks 
contribute to the pairing symmetrically.
The CFL matter is both a superfluid and a color superconductor 
because of the spontaneous breaking of the global $U(1)_{\rm B}$ baryon number
symmetry and the local $SU(3)_{\rm C}$ color symmetry, 
respectively.
It is expected to exist in the cores of dense stars, 
although observational evidence has been elusive.

The purpose of this Letter is to propose 
a possible observational signal of the CFL matter. 
The key ingredients are the topological vortices. 
These vortices are created under rotation owing to the superfluidity of
the CFL matter
\cite{Forbes:2001gj,Iida:2002ev}. 
If the CFL phase is realized in the cores of dense stars,
the creation of vortices is inevitable 
since the stars rotate rapidly.
The superfluid vortices discussed in Refs.~\cite{Forbes:2001gj,Iida:2002ev} 
were found to be dynamically unstable, decaying into sets of constituent vortices \cite{Nakano:2007dr}. 
The stable ones are the so-called
non-Abelian vortices,  
which are superfluid vortices as well as color magnetic flux tubes 
\cite{Balachandran:2005ev}.
Their properties have been studied 
using the Ginzburg-Landau theory 
\cite{Eto:2009kg,
Nakano:2007dr, 
Sedrakian:2008ay, 
Shahabasyan:2009zz, 
Eto:2009bh, 
Eto:2009tr, 
Hirono:2010gq} 
or the Bogoliubov--de Gennes equation \cite{Yasui:2010yw}.
Interestingly, there are fermionic and bosonic degrees of freedom
localized on a vortex.
Non-Abelian vortices are endowed with a novel kind of non-Abelian statistics 
because of the multiple fermion zero modes trapped inside them \cite{Yasui:2010yh}.
On the other hand, the bosonic degrees of freedom are called the
orientational zero modes \cite{Nakano:2007dr,Eto:2009bh,Hirono:2010gq}, 
which are the Nambu-Goldstone bosons that are associated with the symmetry
breaking inside vortices.

\begin{figure}[tbp]
 \begin{center}
  \includegraphics[width=90mm]{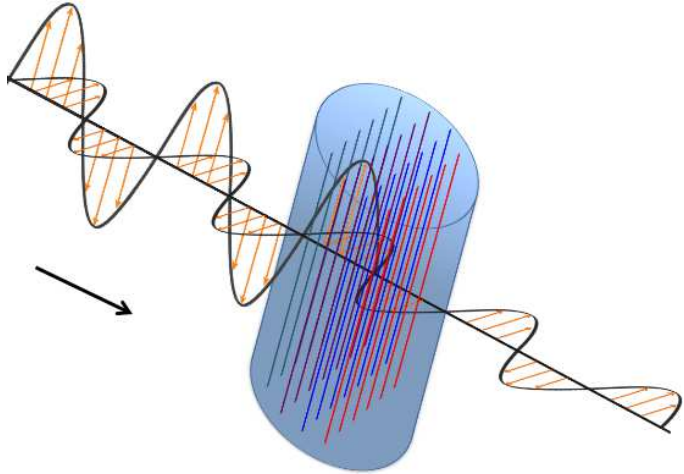}
 \end{center}
 \caption{
Schematic figure of two linearly polarized photons entering a vortex lattice. 
The big arrow represents the propagating direction.
The small arrows indicate the electric field vector.
The waves whose electric fields are parallel to the vortices are
 attenuated, while the ones with perpendicular electric fields are not affected.
}
 \label{fig:polarizer}
\end{figure}

In this Letter, we investigate the electromagnetic properties of
non-Abelian vortices in the CFL phase.
Although the CFL matter itself is electromagnetically neutral,
the orientational zero modes are naturally charged, as is discussed later.
The electromagnetic property of vortices can be 
phenomenologically important 
as it may lead to some observable effects.
As an illustration of such an effect, 
we show that a lattice of vortices in the CFL phase acts as a polarizer of
photons. 
The rotating CFL matter should be threaded with quantum vortices along the
axis of rotation, 
which results in the formation of a vortex lattice 
\cite{Nakano:2007dr,Sedrakian:2008ay,Shahabasyan:2009zz}, 
as in the case of rotating atomic superfluids. 
Suppose that a linearly polarized photon is incident on a vortex lattice (see
Fig.~\ref{fig:polarizer}).
When the electric field of the photon is parallel to the vortices, 
it induces currents along the vortices,
resulting in the attenuation of the photon;
on the other hand, waves with electric fields
perpendicular to the vortices are not affected.
This is exactly what a polarizer does.
A lattice passes electromagnetic waves of a specific polarization and blocks
waves of other polarizations.
This phenomenon, resulting from the electromagnetic interaction of
vortices, 
may be useful in 
finding observational evidence
for the existence of the CFL matter.

In the present analysis, we neglect the mixing of photons and gluons.
The gauge field, $A^\prime_{\mu}$, which remains massless in the CFL phase, 
 is a mixture of the photon $A_{\mu}$ and a part of gluons $A^8_\mu$,
$
 A^\prime_{\mu} = -\sin \zeta A_{\mu} + \cos \zeta A^8_\mu.
$
Here, the mixing angle $\zeta$ is given by 
$
\tan \zeta = \sqrt{3}g/{2e} $\cite{Alford:1998mk}, where $g$ and $e$ are
the strong and electromagnetic coupling constants.
At accessible densities ($\mu \sim 1{\rm GeV}$), the fraction of the
 photon is given by $\sin \zeta \sim 0.999$,
and so, the massless field $A^\prime_\mu$ consists mostly of the ordinary
 photon and includes a small amount of the gluon.
As a first approximation, we neglect the mixing of the gluon to the
massless field.

\emph{Orientational zero modes.}---
The color superconductivity is brought about by the condensation of
diquarks.
At very high densities, the ground state is believed to be the CFL phase, which is
characterized by the spinless and positive parity condensates of the form 
\begin{equation}
 \Phi^a_i = \epsilon_{abc} \epsilon_{ijk} 
\langle 
(q^T)^j_b C \gamma_5 (q)^k_c
\rangle  = \Delta \delta^a_i,
\end{equation}
where $q$ is the quark field, 
$i, j, k =  u,d,s\;(a,b,c =  r,g,b)$ are the flavor (color) indices, $C$ is the charge
conjugation matrix, 
$\Delta$ is a BCS gap function, and the transpose is employed with
respect to the spinor index.
The symmetry breaking pattern is, apart from discrete symmetry,   
\begin{equation}
\begin{split}
 SU(3)_{\rm C} \times SU(3)_{\rm R} \times
SU(3)_{\rm L} \times U(1)_{\rm B} \\
\longrightarrow SU(3)_{\rm C + R + L } \equiv SU(3)_{\rm C + F },
\end{split}
\end{equation}
where $SU(3)_{\rm C}$ is the color symmetry,
$SU(3)_{\rm R(L)}$ is right (left) flavor symmetry, and $U(1)_{\rm B}$ is
the symmetry associated with the baryon number conservation.
The ground state is invariant under the
simultaneous rotation of color and flavor; thus, it is called the
color-flavor locked phase.
In the presence of a vortex, the color-flavor locked symmetry, $SU(3)_{\rm C+F}$, is
further broken down to $SU(2)_{\rm C+F} \times U(1)_{\rm C+F}$ around the
core of the vortex.
Consequently, there appear Nambu-Goldstone (NG) modes confined in the core of
the vortex, which parametrize the coset space
known as the two-dimensional complex projective space
\cite{Nakano:2007dr,Eto:2009bh},
\begin{equation}
 \frac{SU(3)_{\rm C+F}}{SU(2) \times U(1)} \simeq \mathbb{C}P^2.
\end{equation}
There exist classically degenerate vortex solutions, characterized by the
value of $\mathbb{C}P^2$ orientational moduli.
We denote the NG modes by a complex three-component vector $\phi \in
\mathbb{C}P^2$, which satisfies $\phi^\+ \phi=1$.
When we neglect the electromagnetic interaction, 
the low energy effective theory on the vortex which is placed along the
$z$ axis
is shown to be described by the
following $\mathbb{C}P^2$ nonlinear sigma model \cite{Eto:2009bh},
\begin{equation}
\Lcal_{\mathbb{C}P^2} = C \sum_{\alpha = 0, 3} K_\alpha 
\left[
\p^\alpha \phi^\+ \p_\alpha \phi + (\phi^\+ \p^\alpha \phi)  (\phi^\+ \p_\alpha \phi)
\right],
\end{equation}
where the orientational moduli $\phi$ are promoted to dynamical fields, 
and $C$ and $K_\alpha$ are numerical constants.
Under the color-flavor locked transformation, 
the $\mathbb{C}P^2$ fields $\phi$ transform as 
\begin{equation}
 \phi \rightarrow U \phi,
\end{equation}
with $U \in SU(3)_{\rm C+F}$. 

Now, let us consider the electromagnetic fields.
The electromagnetic $U(1)_{\rm EM }$ group is a subgroup of
the flavor group $SU(3)_{\rm F}$, which is generated by $T_8 =
\frac{1}{\sqrt{6}}\diag (-2,1,1)$ in our choice basis.
The electromagnetic interaction is incorporated by gauging the
corresponding symmetry.
Therefore, the low-energy effective action on the vortex should be modified 
to the gauged $\mathbb{C}P^2$ model, 
\begin{equation}
\Lcal_{g\mathbb{C}P^2} = C \sum_{\alpha = 0, 3} K_\alpha 
\left[
\D^\alpha \phi^\+ \D_\alpha \phi + (\phi^\+ \D^\alpha \phi)  (\phi^\+ \D_\alpha \phi)
\right],
\end{equation}
where the covariant derivative is defined by 
\begin{equation}
\D_\alpha \phi = \left( \p_\alpha - ie \sqrt{6} A_\alpha T_8  \right) \phi.
\end{equation}

\emph{Photon-vortex scattering.}---
Here,
we investigate the consequence of the charged degrees of freedom on the vortex.
The low-energy behavior is described by photons propagating in
three-dimensional space and the $\mathbb{C}P^2$ model localized on the
vortex.
Hence, the effective action is given by 
\begin{equation}
S = -\frac{1}{4} \int  F_{\mu\nu}F^{\mu\nu} d^4x
+ 
\int \Lcal_{g \mathbb{C}P^2} dzdt .
\end{equation}

Let us consider the scattering of photons by a vortex.
The equation of motion of the gauge fields derived from 
the effective action is given as 
\begin{equation} 
\begin{split}
\p^\mu F_{\mu\nu}  
&= 
-C K_\nu ie \sqrt{6} 
\ \delta(x_\perp) (\delta_{0\nu} + \delta_{3 \nu} ) \\
 &\times 
\bigl\{
\phi^\+ T_8 \D_\nu \phi  
 - 
(\D_\nu \phi)^\+ T_8 \phi
- 2 \phi^\+ \D_\nu \phi \phi^\+ T_8 \phi 
\bigl\} ,
\end{split}
\end{equation}
where
$\delta ( x_\perp)\equiv  \delta(x)\delta(y)$ is the transverse delta function.
We consider the situation where 
a linearly polarized photon is normally incident on the vortex
and assume that the electric field of the photon is parallel to the
vortex.
Then, the problem is $z$-independent and we can set $\theta=\theta(t)$,
$A_t=A_x=A_y=0$, and $A_z = A_z(t,x,y)$. 
The equation of motion can be rewritten as
\begin{equation}
\begin{split}
 (& \p_t^2  - \p_x^2 - \p_y^2) A_z(t,x,y) \\
 & =
12 C K_3 e^2 
\left\{
\phi^\+ (T_8)^2 \phi + (\phi^\+ T_8 \phi)^2
\right\}
A_z(t,x,y) \ \delta(x_\perp) \\
&\equiv 
12 C K_3 e^2 f(\phi)
A_z(t,x,y) \ \delta(x_\perp), 
\end{split}
\label{eq:gauge-eom}
\end{equation}
where we have defined 
\begin{equation}
f(\phi) \equiv \phi^\+ (T_8)^2 \phi + (\phi^\+ T_8 \phi)^2.
\label{eq:orientation-dep}
\end{equation}
Equation (\ref{eq:gauge-eom}) is the
same as the one treated by Witten in the
context of superconducting strings \cite{Witten:1984eb} except for the
orientation-dependent factor, $f(\phi)$.
The cross section per unit length, $d\sigma/dz$, can be calculated 
by solving the scattering problem, as in Ref.~\cite{Witten:1984eb}, 
\begin{equation}
 \frac{d\sigma}{d z}
=\frac{\left( 12 C K_3 e^2 f(\phi) \right)^2 \eta^2 } {8 \pi} \lambda
= 288 \pi  \left(CK_3 \alpha \eta f(\phi) \right)^2 \lambda,
\end{equation}
where $\lambda$ is the wavelength of the incident photon, $\eta$ is a
numerical factor of order unity, and $\alpha$, the fine structure constant.
On the other hand, if the electric field of the wave is perpendicular to
the vortex, the photon is not scattered since current can flow only
along the vortex.

\emph{Vortex lattice as a polarizer.}---Now let us consider the case
where electromagnetic waves of some intensity normally enter the vortex lattice.
We consider the electric fields of the waves to be parallel to the vortices.
These waves are scattered by the vortices and lose intensity.
The fraction of the loss of intensity when the wave passes through the
lattice for distance $dx$ is
\begin{equation}
\left< \frac{d \sigma}{dz} \right>
n_v dx \equiv \frac{dx}{L},
\end{equation}
where $n_v$ is the number of vortices per unit area. 
Here, $L$ is defined by
\begin{equation}
L \equiv 1/\left( n_v \left< \frac{d\sigma}{dz} \right> \right)=
\ell^2 / \left< \frac{d\sigma}{dz} \right>,
\end{equation}
with the inter-vortex spacing $\ell$. 
As the cross section depends on the internal state (value of $\varphi$) of the vortex,
we have introduced the averaged
scattering cross section $\langle d\sigma/dz \rangle$ over the ensemble
of the vortices.
Let us denote the intensity of waves at distance $x$ from the surface of the
lattice as $I(x)$.
$I(x)$ satisfies 
\begin{equation}
 \frac{I(x+dx)}{I(x)} = 1-\frac{dx}{L}.
\end{equation}
Therefore, the $x$ dependence of $I(x)$ is characterized by the
following differential equation
\begin{equation}
 \frac{I'(x)}{I(x)} = -\frac{1}{L}.
\end{equation}
This equation is immediately solved as 
$
 I(x) = I_0 e^{-x/L},
$
where $I_0$ is the initial intensity. 
Hence, the waves are attenuated with the characteristic length $L$.

We can obtain a rough estimate of the attenuation length.
The total number of vortices can be estimated, as in
Ref.~\cite{Iida:2002ev}, as 
\begin{equation}
 N_v \simeq 1.9 \times 10^{19 } 
\left(\frac{ 1 {\rm ms }}{ P_{\rm rot}} \right)
\left(\frac{\mu/3}{300 {\rm MeV}}  \right)
\left( \frac{R}{10 {\rm km} } \right)^2,
\label{eq:num-vortex}
\end{equation}
where $P_{\rm rot}$ is the rotation period; $\mu$, the baryon
chemical potential; and $R$, the radius of the CFL matter inside dense stars.
These quantities are normalized by typical values.
The intervortex spacing is given by 
\begin{equation}
 \ell \equiv 
\left(\frac{\pi R^2}{N_v}\right)^{1/2} 
\simeq
4.0 \times 10^{-6}~ {\rm m} 
\left(\frac{ P_{\rm rot}}{ 1 {\rm ms }} \right)^{1/2}
\left(\frac{300 {\rm MeV}}{\mu/3}  \right)^{1/2}.
\end{equation}
Therefore, the characteristic decay length of the electromagnetic waves is
roughly estimated as
\begin{equation}
L =
 \frac{\ell^2}{288 \pi  \left(CK_3 \alpha \eta \right)^2 
\left< f(\phi)^2 \right>   \lambda}
\simeq \frac{6.5 \times 10^{-12} \ {\rm m}^2}{\lambda},
\label{eq:decay-length}
\end{equation}
where, we have assumed that the variable $\phi$ is
randomly distributed in the $\mathbb{C}P^2$ space. This assumption is natural
as there is no particularly favored direction in the $\mathbb{C}P^2$
space for the case with three massless flavors \footnote{
The potential quantum mechanically 
induced in the ${\mathbb C}P^2$ model  
is of an exponentially soft scale 
$\Delta \exp[-c (\mu/\Delta)^2]$ 
with the baryon chemical potential $\mu$ \cite{Eto:2011mk}, 
which can be neglected at asymptotically high densities.
}
\footnote{
The presence of a finite strange quark mass does not change the
qualitative feature of the polarizing phoenomenon.
The strange quark mass gives rise to a potential in the effective model,
as discussed in Ref.~\cite{Eto:2009tr}.
When $m_{\rm s}$ is larger than the typical kinetic energy of the
$\mathbb{C}P^2$ modes,  which is given by the
temperature $T \leq T_{\rm c}\sim 10^1 \ {\rm MeV}$, 
and is small enough so that 
the description by the Ginzburg-Landau theory based on the chiral
       symmetry is still valid, 
the orientation of vortices falls into $ \phi_0^T = (0,1,0)$.
This assumption is valid for the realistic value of $m_{\rm
s} \sim 10^2 \ {\rm MeV}$.  
The orientation dependence of the cross section is encapsulated in the
function $f(\phi)$ defined in Eq.~(\ref{eq:orientation-dep}).
Since $f(\phi_0) = 1/3 \neq 0$, photons still interact with the vortex 
in the presence of a finite strange quark mass.
Assuming that all the vortices are with the orientation $\phi_0$, we can
redo the numerical estimates as follows.
The decay length of the photon intensity is recalculated to be 
$
 L \sim (1.2 \times 10^{-11}\ {\rm m}^2) / \lambda, 
$
instead of Eq.~(\ref{eq:decay-length}), and the condition that the
intensity of photons is
significantly decreased within the CFL core of order $1 {\rm km}$ is
given by 
$
 \lambda \geq 1.2 \times 10^{-14} \ {\rm m},
$
instead of Eq.~(\ref{eq:critical-wavelength}).
}.
We have also taken $\eta =1$, $\mu = 900~{\rm MeV}$ and $\Delta = 100
~{\rm MeV} $, from which 
the values of $C$ and $K_3$ are determined accordingly \cite{Eto:2009tr}.
If we adopt the value of $R \sim 1 \ {\rm km}$ for the radius of the CFL core, 
the condition that the intensity is significantly
decreased within the core is written as $L \leq 1 \ {\rm km}$. 
This condition can be rewritten in terms of the wavelength of the photon as
\begin{equation}
\lambda \geq 6.5 \times 10^{-15} \ {\rm m} \equiv \lambda_{\rm c}.
\label{eq:critical-wavelength}
\end{equation}
Therefore, a lattice of vortices serves as a wavelength-dependent filter
of photons.
It filters out the waves with electric fields parallel to the vortices if
the wavelength $\lambda$ is larger than $\lambda_{\rm c}$.
The waves that pass through the lattice are the linearly
polarized ones with the direction of their electric fields perpendicular
to the vortices, as schematically shown in Fig.~\ref{fig:polarizer}.

One may wonder why a vortex lattice with mean vortex distance $\ell$ 
can serve as a polarizer for photons with wavelength many-orders smaller
than $\ell$.
It is true the probability that a photon is scattered during its
propagation for a small distance ($\sim\ell$, for example) is small.
However, while the photon travel through the lattice, the scattering
probability is accumulated and the probability that a photon remains
unscattered decreases exponentially. 
Namely, the small scattering probability is compensated by the large
number of vortices through which a photon passes.
This is why the vortex mean distance and the wavelength of the
attenuated photons can be different.

\emph{Conclusion.}---
We have shown that a quantum vortex in the CFL phase interacts with
photons because of the $\mathbb{C}P^2$ mode on the vortex.
We have demonstrated that, as a consequence, 
photons with electric fields parallel to the vortices 
are attenuated in a vortex lattice.
This effect would be observable if there exist quark stars in which
 the CFL phase continues from the core to the surface. 
However, if the CFL core is covered with a nuclear mantle, 
it would be difficult for optical probes to penetrate the surface of the star.
Even in that case, 
we expect that the electromagnetic properties of vortices
could be useful in finding observational evidence of CFL matter, for
example through the electromagnetic interaction of vortices with strong
magnetic fields in neutron stars.

Y.H. is supported by the
Japan Society for the Promotion of Science for Young Scientists.
M.N. is supported in part by Grants-in-Aid for Scientific Research 
(No. 23103515 and 23740198) from the Ministry of Education, Culture,
Sports, Science and Technology, Japan.

\end{document}